\documentclass[aip,jcp]{revtex4-2}

\usepackage{amsmath}
\usepackage{multirow}

\usepackage{graphicx}
\usepackage{color}

\begin{document}

\newcommand{\red}[1]{{\color{red} #1}}
\newcommand{\blue}[1]{{\color{blue} #1}}

%%%%%%%%%%%%%%%%%%%%%%%%%%%%%%%%%%%%%%%%%%%%%%%%%%%%%%%%%%%%%
% Use the \preprint command to place your local institutional report number 
% on the title page in preprint mode.
% Multiple \preprint commands are allowed.
%\preprint{}

\title{Preparing Small Gaussian Basis for Highly Accurate Ab Initio Description of Lithium Rydberg States} %Title of paper

% repeat the \author .. \affiliation  etc. as needed
% \email, \thanks, \homepage, \altaffiliation all apply to the current author.
% Explanatory text should go in the []'s, 
% actual e-mail address or url should go in the {}'s for \email and \homepage.
% Please use the appropriate macro for the type of information

% \affiliation command applies to all authors since the last \affiliation command. 
% The \affiliation command should follow the other information.

\author{Jan Šmydke}
\email[]{smydke@fzu.cz}
%\homepage[]{Your web page}
%\thanks{}
%\altaffiliation{}
\affiliation{Department of Radiation and Chemical Physics, FZU - Institute of Physics of the Czech Academy of Sciences,
  Na Slovance 1999/2, 18221 Praha 8, Czech Republic}

\author{Benjamín Andreides}
%\email[]{bandreides@gmail.com}
\altaffiliation[also at ]{Institute of Physics, Faculty of Mathematics and Physics, Charles University, Prague, Czech Republic}

\author{Simona Dubcová}
%\email[]{dubcova.simona@gmail.com}

% Collaboration name, if desired (requires use of superscriptaddress option in \documentclass). 
% \noaffiliation is required (may also be used with the \author command).
%\collaboration{}
%\noaffiliation

\date{\today}

%%%%%%%%%%%%%%%%%%%%%%%%%%%%%%%%%%%%%%%%%%%%%%%%%%%%%%%%%%%%%
\begin{abstract}
A new small and highly accurate Gaussian basis set has been developed
for the {\it ab initio} description of $^2$Li Rydberg excited states
up to $n = 7$ and S, P, D, F, and G angular momentum symmetry, and the
appropriate optimization protocol is presented. The obtained Rydberg
excitation energies are compared with results from other highly
accurate approaches. At the \mbox{EOM-CCSD} level of theory the new
basis exhibits higher than $10^{-2}$~eV accuracy of the excitation
energies and the ionization potential and provides superior results
than an analogous universal Gaussian basis set, while utilizing even
smaller number of basis functions. Plots of the 1-dimensional
Rydberg-orbital cuts reveal a regular nodal structure along the
logarithmic scale of the atomic radius reaching tens of {\AA} far from
the nucleus. The presented Rydberg basis set generation methodology is
an important step towards routine {\it ab initio} Rydberg-state
related investigations of more complex systems, such as large atoms
and polyatomic molecules.
\end{abstract}

%\pacs{}% insert suggested PACS numbers in braces on next line

\maketitle %\maketitle must follow title, authors, abstract and \pacs

%%%%%%%%%%%%%%%%%%%%%%%%%%%%%%%%%%%%%%%%%%%%%%%%%%%%%%%%%%%%%
% Body of paper goes here. Use proper sectioning commands. 
% References should be done using the \cite, \ref, and \label commands
\section{Introduction}
\label{Introduction}

% If in two-column mode, this environment will change to single-column format so that long equations can be displayed. 
% Use only when necessary.
%\begin{widetext}
%$$\mbox{put long equation here}$$
%\end{widetext}

% Figures should be put into the text as floats. 
% Use the graphics or graphicx packages (distributed with LaTeX2e).
% See the LaTeX Graphics Companion by Michel Goosens, Sebastian Rahtz, and Frank Mittelbach for examples. 
%
% Here is an example of the general form of a figure:
% Fill in the caption in the braces of the \caption{} command. 
% Put the label that you will use with \ref{} command in the braces of the \label{} command.
%
% \begin{figure}
% \includegraphics{}%
% \caption{\label{}}%
% \end{figure}

% Tables may be be put in the text as floats.
% Here is an example of the general form of a table:
% Fill in the caption in the braces of the \caption{} command. Put the label
% that you will use with \ref{} command in the braces of the \label{} command.
% Insert the column specifiers (l, r, c, d, etc.) in the empty braces of the
% \begin{tabular}{} command.
%
% \begin{table}
% \caption{\label{} }
% \begin{tabular}{}
% \end{tabular}
% \end{table}

Rydberg states of atoms and molecules are characterized by a
predominant excitation of a single electron to a high energy orbit. In
contrast to valence excited states, where motion of all the electrons
is confined to a few {\AA} around nuclei, the Rydberg excited electron
moves rather tens, hundreds and thousands {\AA} far from the remaining
ion core. From such enormous distances, the ion core appears to the
Rydberg electron like a point charge that makes it orbit similarly
like in the hydrogen atom. This resemblance is pronounced in an
analogous structure of the Rydberg states energy spectrum with an
infinite number of levels approaching a distinct ionization
threshold. From this perspective, the Rydberg states are also commonly
viewed as states of an electron captured by the ion core with a
specific electron affinity.

Correct {\it ab initio} description of Rydberg states is highly
important in computational chemistry since they directly affect the
shapes of potential energy surfaces (PES) due to common interactions
with the valence states of similar energy.~\cite{2009ReKr} If the
Rydberg states cannot be sufficiently described by the utilized basis
sets, or are omitted from the configurational space, the related
chemical reactivity and spectroscopic calculations may have serious
convergence problems or lead to incorrect results.

Compared to a routine {\it ab initio} treatment of valence excited
states, an {\it ab initio} modeling of Rydberg states has always been
challenging due to their extremely diffuse character that requires
utilization of appropriately long range basis sets. For molecules, in
particular, common quantum chemistry codes use exclusively only
Gaussian-type basis sets, which quickly dissipate, effectively within
several {\AA}, and, therefore, are primarily suitable for systems
where all electrons are strongly attracted to nuclei, like in cations
or neutral ground and valence excited states.

To model electronic motion in large distances, one can extend a
standard Gaussian basis by a series of diffuse, {\it i.e.}  low
exponent Gaussians. However, if such an augmented basis is not
specifically optimized, it may result in a rather qualitative
description of merely one or two Rydberg states. In addition, the
extra diffuse functions often cause near linear dependence within the
basis, which necessarily spoils or at least strongly affects the
numerical computations. The ability to obtain reliable Gaussian basis
sets specific for the description of Rydberg states would thus mean a
great benefit for computational chemistry.

Kaufmann~{\it et al.}~\cite{1989KaBaJu} developed a systematic
approach for generating primitive Gaussian basis sets suitable for
modeling either the atomic Rydberg or the lowest continuum
states. The approach is based on a one-to-one mapping of Gaussians to
appropriate exponential functions that span a complete bound or
continuum atomic basis. It provides easy to use formulas and
coefficients for generating atomic Gaussian exponents that are,
however, system universal, and as such may be suboptimal, and quickly
become too large to handle.

In our recent work~\cite{2023Sm} we demonstrated how a tailored
Rydberg basis could be constructed by utilizing a simple variational
algorithm. Accordingly optimized Gaussian basis then allowed an {\it
  ab initio} computation of a large number of highly accurate Rydberg
states of a lithium atom. To our knowledge, the high number of lithium
Rydberg states computed at the EOM-CCSD level has so far not been
achieved by other {\it ab initio} approaches. Only the few lowest
states were also known from the theoretically most accurate, but
extremely demanding computations, employing the Explicitly Correlated
Gaussians (ECG) or performing the Full-CI on optimized
Slater-type-orbitals (STO), and the results were found fully
competitive.

Despite the undoubtedly promising results achieved by the relatively
cheap variational method, the high number of primitive Gaussians
needed to form the optimal contracted functions prevented the
developed basis from utilization by common quantum chemistry codes,
which often have hard-coded restrictions on basis sets. Nevertheless,
the methodology had still plenty of room for improvement in order to
reduce the final basis set size. Moreover, modeling of a particular
chemical system may not always require accurate description of dozens
of Rydberg states, but my need to obtain just a few of the lowest
states. Accordingly, computing only the lowest and well-separated
energy levels could also allow for relaxing the numerical precision.

In this paper, we introduce a new Gaussian basis for lithium, optimized
for Rydberg excited states with the principal quantum number $n$
ranging from 2 only to 7 while sticking to the angular momentum
symmetry range from S to G. The new basis utilizes a more suitable
parameterization, requires only a moderate number of primitive
Gaussians within the contracted functions, and is capable of
Rydberg-state-selective computations with only a few added Rydberg
functions while preserving the high accuracy.

In the earlier study~\cite{2023Sm} we focused mainly on the proof of
the concept, and demonstrated how many and how accurately the Rydberg
states could be calculated using merely a Gaussian-type basis and
standard methods of quantum chemistry. Only later we looked into the
actual shapes of the Rydberg orbitals and discovered their very
regular nodal structure reaching from the vicinity of the nucleus all
the way up to almost $100,000$ {\AA} far. The plots of the basis
functions also revealed a source of a near linear dependence that
affected the basis set optimization and subsequently enabled us to
resolve this issue when designing the new basis. For its valuable
insight, the nodal structure of the Rydberg orbitals, obtained from
the new as well as from the earlier developed basis,~\cite{2023Sm} is
discussed in this study.

The rest of the article is organized as
follows. Section~\ref{Methodology} summarizes the principle behind the
tailored Rydberg basis optimization, specifies a suitable
parameterization for the Gaussian basis, and describes the practical
optimization protocol that led to the final optimal $^2$Li Gaussian
Rydberg basis set. The necessary computational details are also
specified. Results are presented and discussed in
Section~\ref{Results}, where the relevant basis sets are first
introduced and named. Then the computed excitation energies,
ionization potentials, and the nodal structure of the Rydberg orbitals
are thoroughfully discussed. The conclusions are finally drawn in
Section~\ref{Conclusions}. The Supporting Information (SI) includes
all the plots of the Rydberg orbitals generated utilizing the new as
well as the original basis set. The developed basis is also included
as a formatted GENBAS file.

%%%%%%%%%%%%%%%%%%%%%%%%%%%%%%%%%%%%%%%%%%%%%%%%%%%%%%%%%%%%%
\section{Methodology}
\label{Methodology}

\subsection{Approximating Rydberg Orbitals}
\label{Principle}
The idea behind the tailored Rydberg basis construction~\cite{2023Sm}
lies in the Koopmans' theorem,~\cite{SzaboOstlund_1996} which
approximates values of the system's electron affinity with the energy
of its standard canonical virtual orbitals taken with an opposite sign

\begin{equation}
  \label{EqKoopmans}
  {\rm EA}_a \approx - \varepsilon_a
\end{equation}
where ${\rm EA}_a$ is the approximate electron affinity when capturing
an electron in the $a$-th virtual orbital and $\varepsilon_a$ is the
orbital's energy. In other words, the standard canonical virtual
orbitals with negative energies are able to capture electrons while
forming bound states. For a field generated by a given Hartree--Fock
wave function, the virtual orbitals can be variationally optimized by
minimizing the orbital energies while tuning parameters of additional
(diffuse) basis. For systems with Rydberg states, the process will
result in a set of Rydberg orbitals, which can eventually serve as the
desired contracted Rydberg basis functions. It should be emphasized
that the electron excited to a Rydberg orbital (corresponding to a
Rydberg state) describes actually an electron captured by the ion
core. Thus, for example, to obtain Rydberg orbitals of a neutral
$^2$Li atom, one has to optimize virtual orbitals of the $^1$Li$^+$
cation.

Since the Rydberg electron prevails mostly very far from the ion core,
its correlation with the other electrons is weak. Hence, the
Hartree--Fock approximation of the Rydberg orbitals can be reasonably
accurate. Nevertheless, it needs to be stressed that the Koopmans'
theorem is not implicitly valid for other than the canonical
restricted closed shell Hartree--Fock theory. For the unrestricted
(UHF) or restricted open shell (ROHF) Hartree--Fock approaches, the
Koopmans' theorem validity has to be explicitly
guaranteed.~\cite{2006PlGoBr,2010DaPl,2014PlDa,2018Pl} It should also
be noted that the described principle allows application to arbitrary
basis set types, which can, of course, be just Gaussians.

\subsection{Composition of the Gaussian Rydberg Basis}
\label{GaussianBasisComposition}
In our earlier work~\cite{2023Sm} we developed a large Gaussian basis
for lithium Rydberg states (here referred to as 2023Sm) that consisted
of three components. The standard \mbox{aug-ANO-pVQZ}
basis,~\cite{2011NeVa}, which assured a sufficient description of the
electronic correlation, an additional optimized contracted S function
that practically helped keeping the Hartree--Fock energy of the Li$^+$
ion core invariant to the varying diffuse basis set parameters, and
finally the contracted S, P, \dots, G optimized Rydberg functions. The
primitive Gaussians used in the contracted additional S and in each of
the Rydberg angular momentum subspaces used the even tempered Gaussian
(ETG) prescription for the exponents

\begin{equation}
  \label{EqETG}
  \zeta_k \, = \, \zeta_0 \; \alpha^{k};\;\;\;
  \alpha \in (0,1),\;
  k = 0 \, \dots \, (N-1)
\end{equation}
% or
%\begin{equation}
%  \label{EqETGLog}
%  log \, \zeta_k \, = \, log \, \zeta_0 \; + \; k\;log \, \alpha;\;\;\;
%  log \, \alpha < 0,\;
%  k = 0 \, \dots \, (N-1)
%\end{equation}
that offered a bare minimum number of variational parameters.

In the present study, to improve flexibility of the primitive
Gaussians, we turned to an analogous formula that we developed a
decade ago for a description of Rydberg and resonance states of
helium~\cite{2013KaSm,2013KaSmCi}, and which we called as the
exponentially tempered Gaussians (ExTG)

\begin{equation}
  \label{EqExTG}
  \zeta_k \, = \, \zeta_0 \; \alpha^{[(e^{\beta k} - 1) / \beta]};\;\;\;
  \alpha \in (0,1),\;
  \beta < 0,\;
  k = 0 \, \dots \, (N-1)
\end{equation}
or, logarithmically,
\begin{equation}
  \label{EqExTGLog}
  log \, \zeta_k \, = \, log \, \zeta_0 \; + \; \frac{(e^{\beta k} - 1)}{\beta} \; log \, \alpha;\;\;\;
  log \, \alpha < 0,\;
  \beta < 0,\;
  k = 0 \, \dots \, (N-1)
\end{equation}
Either forms allow even more flexibility via an expansion

\begin{equation}
  \label{EqGenExTGA}
  \zeta_k \, = \, \sum_i \; \zeta_{0,i} \; \alpha_{i}^{[(e^{\beta_{i} k} - 1) / \beta_{i}]};\;\;\;
  \alpha_i \in (0,1),\;
  \beta_i < 0,\;
  k = 0 \, \dots \, (N-1)
\end{equation}
or
\begin{equation}
  \label{EqGenExTGB}
  log \, \zeta_k \, = \, log \, \zeta_0 \; + \; \sum_i \; \frac{(e^{\beta_{i} k} - 1)}{\beta_{i}} \; log \, \alpha_{i};\;\;\;
  log \, \alpha_i < 0,\;
  \beta_i < 0,\;
  k = 0 \, \dots \, (N-1)
\end{equation}
where the latter is slightly less flexible than the former.  It was
found that both formulas led to a similar quality of the generated
basis sets with a sufficient expansion into only two terms, {\it
  \mbox{i. e.}} $i \in \{1, 2\}$. The results presented in this study
were achieved with the ExTG series variant given by
Eq.~\ref{EqGenExTGA}. The basis developed in this study contains only
the \mbox{aug-ANO-pVQZ} set extended with the contracted optimal
Rydberg functions. The extra S function that kept the Hartree--Fock
energy at the limit during the optimization process had only
negligible effect on the correlated excited states, so it was
eventually excluded.

\subsection{The Optimization Protocol}
\label{Protocol}
To obtain the Gaussian basis tailored for $^2$Li Rydberg states (up to
$n=7$), the following optimization protocol has been suggested.

\begin{enumerate}
  \item Select a standard basis of a moderate size, capable of a
    sufficient electronic correlation description. The choice is
    \mbox{aug-ANO-pVQZ}~\cite{2011NeVa} and will be used in all
    computations.

\item Achieve the $^1$Li$^+$ Hartree--Fock energy limit ($E_{\rm
  HF}^{\rm lim}$) by tuning a primitive S Gaussian series given by
  Eq.~(\ref{EqGenExTGA}) that is added to the \mbox{aug-ANO-pVQZ}
  basis, to make sure that the Hartree--Fock field will not vary when
  more basis functions are inserted in the subsequent Rydberg orbital
  optimization step. Once achieved, contract the ExTG S primitives
  from the 1S LCAO coefficients and add the resulting contracted
  function to the \mbox{aug-ANO-pVQZ} basis for all further
  computations.

\item Optimize the first S Rydberg orbital of the $^1$Li$^+$ system,
  {\it \mbox{i. e.}} 2S (actually a valence orbital for the neutral
  $^2$Li system) by minimizing its energy and tuning a new S ExTG
  series (that is added to the \mbox{aug-ANO-pVQZ} together with the
  $E_{\rm HF}^{\rm lim}$ optimized S function), while making sure the
  Hartree--Fock energy is kept at $E_{\rm HF}^{\rm lim}$. Once the
  orbital is optimal, continue optimizing higher Rydberg S orbitals,
  too, one by one, up to 7S, by further tuning the S ExTG series and
  making sure all the previously optimized Rydberg S orbital energies
  together with the Hartree--Fock energy are kept constant. Finally,
  contract the ExTG primitives from the LCAO coefficients of the
  resulting Rydberg S orbitals.

\item Perform the third step independently also for all the other desired
  angular momenta (P, D, F and G) and store the resulting contracted
  Rydberg functions.
\end{enumerate}

\subsection{Computational Details}
\label{computational}

All the Hartree--Fock and orbital energies necessary for the basis set
optimization were evaluated by the MRCC~\cite{mrcc_orig,mrcc_2020}
quantum chemistry package. The program mdoptcli~\cite{mdoptcli} was
utilized for the multivariate energy minimization, where the
convergence criterion for the Rydberg orbital energies was set to
$10^{-9}$~a.~u. The computations involving electronic correlation for
either the ground or the excited states (RHF/ROHF CCSD and ROHF
EOM-CCSD) were performed by the program GAMESS~\cite{gamess_2020}.

%%%%%%%%%%%%%%%%%%%%%%%%%%%%%%%%%%%%%%%%%%%%%%%%%%%%%%%%%%%%%
\section{Results and Discussion}
\label{Results}

\subsection{Optimized Rydberg Basis Sets}
Before presenting the $^2$Li Rydberg states results, let us first
briefly summarize the basis sets used in this study.

As noted earlier, the basis set obtained by the protocol introduced in
Sect.~\ref{Protocol} could eventually drop the extra S function that
was used for keeping the Hartree--Fock energy at its limit during the
optimization of the Rydberg orbitals, since the subsequent correlated
computations revealed it had a negligible effect on the excitation
energies. In this manuscript we call the final basis as
\mbox{AAQZ-ExTG-Ry7G}, meaning that it is based on the standard
\mbox{aug-ANO-pVQZ} basis with added Rydberg functions capable of
description of all the lowest Rydberg states from 2S up to 7G and
which were constructed from ExTG primitives.

The basis could be further reduced to a bare minimum
\mbox{AAQZ-ExTG-Ry3P}~+~nL that allows a state selective computation
of higher than 3P Rydberg states, just by adding a single particular
Rydberg basis function ({\it e. g.} 5D or 4F), resulting in a
significantly smaller utilized basis set, while maintaining the
accuracy of the full \mbox{AAQZ-ExTG-Ry7G}.

Our earlier developed lithium basis~\cite{2023Sm} was also based on
the \mbox{aug-ANO-pVQZ}, but with the Rydberg functions formed from
ETG primitives. To clearly distinguish the basis sets, we name the
original large ETG-based basis simply as 2023Sm, as a reference to the
publication. Its minimal variant for state selective computations is
denoted as 2023Sm-Ry7D, which also reveals how incomparably larger it
is than the newly developed \mbox{AAQZ-ExTG-Ry3P}.

For a direct comparison of our developed Rydberg basis sets with the
well established universal primitive Gaussian basis~\cite{1989KaBaJu}
we constructed a comparable basis set that should describe all the
Rydberg states up to $n=7$ with appropriate Gaussian primitives added
to the \mbox{aug-ANO-pVQZ}. The basis is denoted as
AAQZ-1989KaBaJu-Ry7G.

In Table~\ref{TabBasisSize} one can compare numbers of primitive and
contracted basis functions contained in all the respective basis
sets. It can be clearly seen that the number of primitives in the
newly developed ExTG-based sets is cut by almost one half compared to
the 2023Sm basis. Even though the \mbox{AAQZ-ExTG-Ry7G} was optimized
for Rydberg states only up to $n=7$, the optimization process
confirmed the superiority of ExTG over ETG not only in the smaller
number of Gaussian primitives necessary for a given number of optimal
Rydberg orbitals, but also in smoother optimization process with much
less near linear dependence issues.

The most important difference is, however, in the size of the minimal
contracted state selective basis sets. The new \mbox{AAQZ-ExTG-Ry3P}
requires addition of only two S, and two P Rydberg functions, {\it
  \mbox{i. e.}} 8 distinct functions to the standard aug-ANO-pVQZ
basis, where the 2023Sm-Ry7D needed eight S, six P and five D
additional Rydberg functions, making altogether 51 extra functions to
the standard aug-ANO-pVQZ. Considering the unfavourable scaling of the
electron correlation methods on the size of the virtual orbital space,
the basis size reduction achieved in the \mbox{AAQZ-ExTG-Ry3P} basis
means a significant success.

The Rydberg functions in the AAQZ-1989KaBaJu-Ry7G basis are
represented by pairs of Gaussian primitives,~\cite{1989KaBaJu} so the
basis is larger than the tailored contracted \mbox{AAQZ-ExTG-Ry7G},
although each contracted function consists of several
primitives. Nevertheless, as shown in the following subsection, the
Rydberg energy levels computed with the tailored basis sets are
superior to the results from the universal Gaussian basis. Not only
may this suggest that the tailored basis set optimization is
favourable to employing universal basis sets, but also that the
exponentially tempered Gaussian series is highly appropriate for the
Rydberg orbital description.

%%% TabBasisSize
\begin{table}
  \caption{\label{TabBasisSize} Numbers of primitive and contracted Gaussian basis functions in the discussed basis sets.}
  \begin{tabular}{c|c|c|c|c|c|c}
    \hline \hline
    Basis & type & S & P & D & F & G \\
    \hline \hline
    \multirow{2}{*}{\mbox{aug-ANO-pVQZ}} & prim. & 19 & 8 & 3 & 2 & \\
    & contr. & 7  & 4 & 3 & 2 & \\
    \hline \hline
    \multirow{2}{*}{2023Sm-Ry7D$^*$} & prim. & 19 30 & 50 & 40 & 30 & 25 \\
    & contr. & 1  7 & 6 & 5 & & \\
    \hline
    \multirow{2}{*}{\mbox{AAQZ-ExTG-Ry7G}$^*$} & prim. & 23 & 28 & 24 & 16 & 18 \\
    & contr. & 6 & 6 & 5 & 4 & 3 \\
    \hline
    \multirow{2}{*}{\mbox{AAQZ-ExTG-Ry3P}$^*$} & prim. & 23 & 28 & 24 & 16 & 18 \\
    & contr. & 2 & 2 & & & \\
    \hline
    AAQZ-1989KaBaJu-Ry7G$^*$ & prim. & 12 & 12 & 10 & 8 & 6 \\
    \hline \hline
    \multicolumn{7}{l}{$^*$the numbers exclude the \mbox{aug-ANO-pVQZ} basis functions}
  \end{tabular}
\end{table}

%%%%%%%%%%%%%%%%%%%%%%%%%%%%%%%%%%%%%%%%%%%%%%%%%%%%%%%%%%%%%
\subsection{Rydberg Excitation Energies and the Ionization Limit}
\label{Energies}

The computed ROHF/EOM-CCSD excitation energies from the ground 2S
state of the neutral $^2$Li to all the Rydberg states up to $n=7$ and
in angular momentum symmetry S to G are presented in
Table~\ref{TabEE}. Each column corresponds to the given basis set,
where the state-selective results in the column B were computed using
the \mbox{AAQZ-ExTG-Ry3P} basis with the appropriately added nL
Rydberg basis function ({\it e. g.} for computing the 4D Rydberg
state, the 4D Rydberg basis function was added). Results from the
earlier developed 2023Sm basis are in column C, while in column D the
AAQZ-1989KaBaJu-Ry7G universal Gaussian basis was utilized. The
reference values in the last column (E), if available, represent the
non-relativistic excitation energies achieved by methods at the
highest level of theory, most of them also assigned as the exact
non-relativistic estimates. For easier comparison of the different
basis sets, Table~\ref{TabEEDif} conveniently shows differences of the
respective columns.

%%% TabEE
\begin{table}
  \caption{\label{TabEE} Comparison of the computed \mbox{EOM-CCSD}
    excitation energies of $^2$Li in various basis sets with the
    non-relativistic results at the highest level of theory. Values
    corresponding to the exact non-relativistic estimate are also
    marked with an asterisk. The values are in eV.}
  \begin{tabular}{c|c|c|c|c|c}
    \hline \hline
    \multirow{2}{*}{State} & A         & B              & C      & D                    & E         \\
    \cline{2-6}
          & \mbox{AAQZ-ExTG-Ry7G} & \mbox{AAQZ-ExTG-Ry3P} + nL & 2023Sm~\cite{2023Sm} & \mbox{AAQZ-1989KaBaJu-Ry7G} & reference \\
    \hline \hline
    3S    & 3.3704    & 3.3704         & 3.3704 & 3.3708               & 3.3732$^*$~\cite{2008PuPa,2019BrBuStAd} \\
    4S    & 4.3378    & 4.3378         & 4.3378 & 4.3381               & 4.3410$^*$~\cite{2009SiHa,2019BrBuStAd} \\
    5S    & 4.7452    & 4.7452         & 4.7456 & 4.7454               & 4.7486$^*$~\cite{2009SiHa,2019BrBuStAd} \\
    6S    & 4.9545    & 4.9544         & 4.9561 & 4.9546               & 4.9579$^*$~\cite{2009SiHa,2019BrBuStAd} \\
    7S    & 5.0760    & 5.0760         & 5.0784 & 5.0761               & 5.0795$^*$~\cite{2009SiHa,2019BrBuStAd} \\
    \hline
    2P    & 1.8472    & 1.8473         & 1.8472 & 1.8910               & 1.8478$^*$~\cite{2012WaYaQiDr,2012BuAd} \\
    3P    & 3.8318    & 3.8319         & 3.8317 & 3.8493               & 3.8343~\cite{2012BuAd} \\
    4P    & 4.5186    & 4.5187         & 4.5186 & 4.5267               & 4.5217~\cite{2012BuAd} \\
    5P    & 4.8341    & 4.8341         & 4.8341 & 4.8384               & 4.8374~\cite{2012BuAd} \\
    6P    & 5.0046    & 5.0046         & 5.0046 & 5.0071               & 5.0080~\cite{2012BuAd} \\
    7P    & 5.1070    & 5.1070         & 5.1070 & 5.1086               & 5.1104~\cite{2012BuAd} \\
    \hline
    3D    & 3.8753    & 3.8754         & 3.8754 & 3.9072               & 3.8786$^*$~\cite{2012WaYaQiDr,2011ShBuAd} \\
    4D    & 4.5374    & 4.5373         & 4.5373 & 4.5570               & 4.5408~\cite{2011ShBuAd} \\
    5D    & 4.8437    & 4.8437         & 4.8437 & 4.8554               & 4.8472~\cite{2011ShBuAd} \\
    6D    & 5.0102    & 5.0102         & 5.0101 & 5.0174               & 5.0137~\cite{2011ShBuAd} \\
    7D    & 5.1105    & 5.1105         & 5.1088 & 5.1154               & 5.1140~\cite{2011ShBuAd} \\
    \hline
    4F    & 4.5379    & 4.5379         & 4.5379 & 4.5617               & 4.5414$^*$~\cite{1999Ki} \\
    5F    & 4.8440    & 4.8440         & 4.8440 & 4.8629               & 4.8475$^*$~\cite{1999Ki} \\
    6F    & 5.0104    & 5.0103         & 5.0103 & 5.0235               &        \\
    7F    & 5.1106    & 5.1106         & 5.1106 & 5.1197               &        \\
    \hline
    5G    & 4.8441    & 4.8441         & 4.8437 & 4.8555               & 4.8371~\cite{2013RuMaFr} \\
    6G    & 5.0104    & 5.0103         & 5.0104 & 5.0215               & 5.0041~\cite{2013RuMaFr} \\
    7G    & 5.1106    & 5.1106         & 5.1106 & 5.1209               & 5.1045~\cite{2013RuMaFr} \\
    \hline \hline
  \end{tabular}
\end{table}

%%% TabEEDif
\begin{table}
  \caption{\label{TabEEDif} Differences between various columns of Table~\ref{TabEE}. The values are in eV.}
  \begin{tabular}{c|c|c|c|c|c}
    \hline \hline
    State & $\rm B - \rm A$ & $\rm A - \rm E$ & $\rm B - \rm E$ & $\rm C - \rm E$ & $\rm D - \rm E$ \\
    \hline \hline
    3S & -1.0E-05 & -2.7E-03 & -2.7E-03 & -2.7E-03 & -2.4E-03 \\
    4S & -2.8E-06 & -3.2E-03 & -3.2E-03 & -3.2E-03 & -3.0E-03 \\
    5S & -3.1E-05 & -3.4E-03 & -3.4E-03 & -3.0E-03 & -3.2E-03 \\
    6S & -3.4E-05 & -3.4E-03 & -3.5E-03 & -1.8E-03 & -3.3E-03 \\
    7S & -3.9E-05 & -3.5E-03 & -3.5E-03 & -1.1E-03 & -3.4E-03 \\
    \hline
    2P &  9.2E-05 & -5.2E-04 & -4.3E-04 & -5.3E-04 &  4.3E-02 \\
    3P &  1.1E-04 & -2.6E-03 & -2.5E-03 & -2.6E-03 &  1.5E-02 \\
    4P &  7.7E-05 & -3.1E-03 & -3.0E-03 & -3.1E-03 &  5.0E-03 \\
    5P & -1.7E-05 & -3.3E-03 & -3.3E-03 & -3.3E-03 &  1.0E-03 \\
    6P & -3.4E-05 & -3.4E-03 & -3.4E-03 & -3.4E-03 & -8.7E-04 \\
    7P & -4.6E-05 & -3.4E-03 & -3.5E-03 & -3.4E-03 & -1.8E-03 \\
    \hline
    3D &  7.5E-05 & -3.3E-03 & -3.2E-03 & -3.2E-03 &  2.9E-02 \\
    4D & -9.3E-06 & -3.4E-03 & -3.4E-03 & -3.4E-03 &  1.6E-02 \\
    5D & -1.7E-05 & -3.5E-03 & -3.5E-03 & -3.5E-03 &  8.2E-03 \\
    6D & -1.8E-05 & -3.5E-03 & -3.5E-03 & -3.6E-03 &  3.8E-03 \\
    7D & -2.8E-05 & -3.5E-03 & -3.5E-03 & -5.3E-03 &  1.3E-03 \\
    \hline
    4F &  2.6E-05 & -3.5E-03 & -3.5E-03 & -3.5E-03 &  2.0E-02 \\
    5F & -1.6E-06 & -3.5E-03 & -3.5E-03 & -3.5E-03 &  1.5E-02 \\
    6F & -1.0E-05 &          &          &          &          \\
    7F & -1.3E-05 &          &          &          &          \\
    \hline
    5G & -1.4E-05 &  6.9E-03 &  6.9E-03 &  6.6E-03 &  1.8E-02 \\
    6G & -1.6E-05 &  6.3E-03 &  6.2E-03 &  6.2E-03 &  1.7E-02 \\
    7G & -1.8E-05 &  6.1E-03 &  6.1E-03 &  6.1E-03 &  1.6E-02 \\
    \hline \hline
  \end{tabular}
\end{table}

Let us first look into the newly developed \mbox{AAQZ-ExTG-Ry7G} basis
and compare it with the old 2023Sm basis results. To see how their
excitation energies differ from the reference values (let us call the
difference from the reference value as an error), we compare columns
$\rm A - \rm E$ and $\rm C - \rm E$ in Table~\ref{TabEEDif}. One can
see both basis sets exhibit similar errors with a trend converging
towards -3.5 meV for higher $n$, except for the G states, where the
errors converge towards 6~meV. All the states have practically same
errors in one basis as in the other, except for the states 5S--7S and
7D, in which the new basis set keeps the trend -3.5~meV error value,
while the 2023Sm basis deviates. Conforming to the trend might suggest
the \mbox{AAQZ-ExTG-Ry7G} basis may be superior to the 2023Sm for the
states it has been developed, though no justification for the observed
trend is given.

Second, we focus on the minimal \mbox{AAQZ-ExTG-Ry3P}~+~nL basis and
the quality of its excitation energies by comparing them to the full
\mbox{AAQZ-ExTG-Ry7G} basis set results, which is shown in the $\rm B
- \rm A$ column of Table~\ref{TabEEDif}. We see the results are almost
identical, with the largest difference between the two basis sets as
0.1 meV for the state 3P, while for the other states differing mostly
by one order or even two orders of magnitude less.

Third, we compare the excitation energy errors between the
\mbox{AAQZ-ExTG-Ry7G} and the universal AAQZ-1989KaBaJu-Ry7G basis
that are shown in columns $\rm A - \rm E$ and $\rm D - \rm E$ of
Table~\ref{TabEEDif}. While the results for the S Rydberg states are
very similar, for the other angular momenta the AAQZ-1989KaBaJu-Ry7G
basis cannot match with the accuracy of the tailored basis and the
errors of the adjacent energy levels rather fluctuate than follow a
trend.

%%% TabIP
\begin{table}
  \caption{\label{TabIP} Comparing different basis sets on the
    computed ionization potential at the CCSD level of theory,
    evaluated as ${\rm IP} = E(^1{\rm Li}^{+}) - E(^2{\rm Li})$.}
  \begin{tabular}{l|r}
    \hline \hline
    Basis & IP / eV \\
    \hline \hline
    \mbox{AAQZ-ExTG-Ry7G} & 5.38766 \\
    \mbox{AAQZ-ExTG-Ry3P} & 5.38764 \\
    2023Sm    & 5.38766 \\
    AAQZ-1989KaBaJu-Ry7G & 5.38766 \\
    \hline
    experiment~\cite{Lide_1992} & 5.39172 \\
    \hline \hline
  \end{tabular}
\end{table}

Finally, let us see how the discussed basis sets predict the value of
the ionization potential (IP), which should make the ultimate limit of
the Rydberg energy level series. The IP has been computed as the
difference between the ground state CCSD energy of the $^1$Li$^+$
cation and the $^2$Li neutral atom and the results are shown in
Tab.~\ref{TabIP}. All the basis sets performed very well, matching the
value of the large 2023Sm basis~\cite{2023Sm}. To spot a difference,
we provide one more decimal digit to reveal only negligible deficiency
in the minimal \mbox{AAQZ-ExTG-Ry3P} state-selective basis. The
computed IP is only 4~meV below the experimental value, perfectly
matching the error of the excitation energies.  We could thus conclude
the accuracy of the \mbox{EOM-CCSD} Rydberg energy levels achievable
by the newly developed basis sets \mbox{AAQZ-ExTG-Ry7G} and
\mbox{AAQZ-ExTG-Ry3P}~+~nL is a few meV, {\it i. e.} higher than
$10^{-2}$~eV.

%%%%%%%%%%%%%%%%%%%%%%%%%%%%%%%%%%%%%%%%%%%%%%%%%%%%%%%%%%%%%
\subsection{Rydberg Orbitals}
\label{Plots}

The Rydberg orbitals have been computed as the canonical RHF virtual
orbitals of the $^1$Li$^+$ cation that have negative orbital
energy. The orbitals thus approximate an electron captured by the
$^1$Li$^+$ cation, which corresponds to the Rydberg excited electron
in the neutral $^2$Li system. This way, the 2S orbital can also be
considered as Rydberg, even though it is a valence orbital for the
$^2$Li.

In this section we discuss the nodal structure of the Rydberg
orbitals, which can be conveniently depicted with one-dimensional
orbital cuts along the $z$~axis. When the orbitals are degenerate (P,
D, F, or G), the orbital with the $z^l$ angular component is selected
({\em \mbox{i. e.}}  ${\rm P}_z$, ${\rm D}_{z^2}$, \dots) and only the
appropriate radial nodal structure is shown. The Supporting
Information (SI) contains plots of all the relevant orbitals computed
in the \mbox{aug-ANO-pVQZ}, \mbox{AAQZ-ExTG-Ry7G}, and 2023Sm basis
sets. Here only the most important cases are discussed.

\begin{figure}[htb]
\includegraphics[height=0.88\textheight]{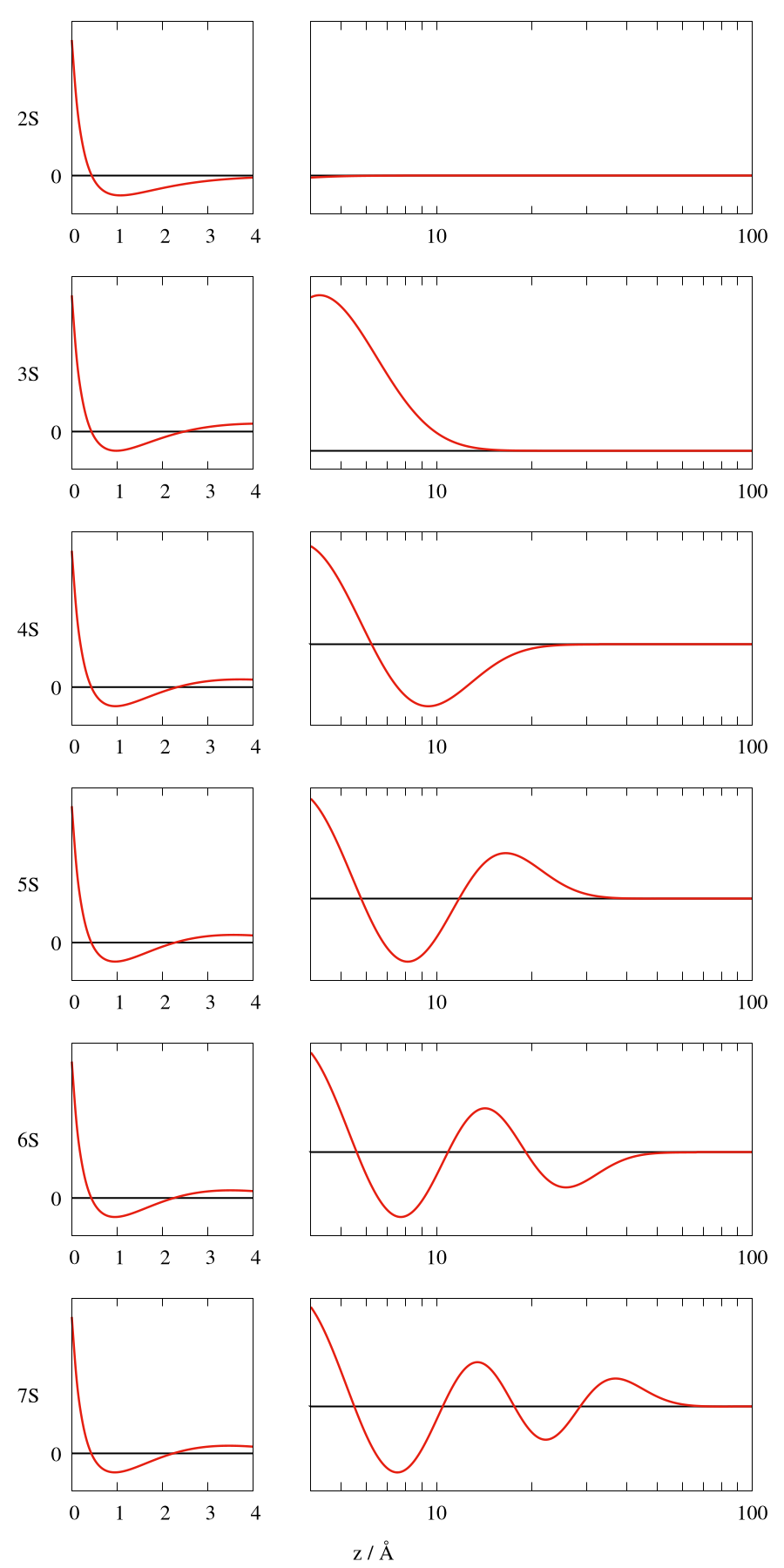}
\caption{\label{FigS}Rydberg S orbitals of the $^2$Li as 1-dimensional
  cuts along the $z$~axis, and obtained as the RHF virtual orbitals of
  the $^1$Li$^+$ cation with negative orbital energy. Computed in the
  \mbox{AAQZ-ExTG-Ry7G} basis.}
\end{figure}

\begin{figure}[htb]
\includegraphics[height=0.88\textheight]{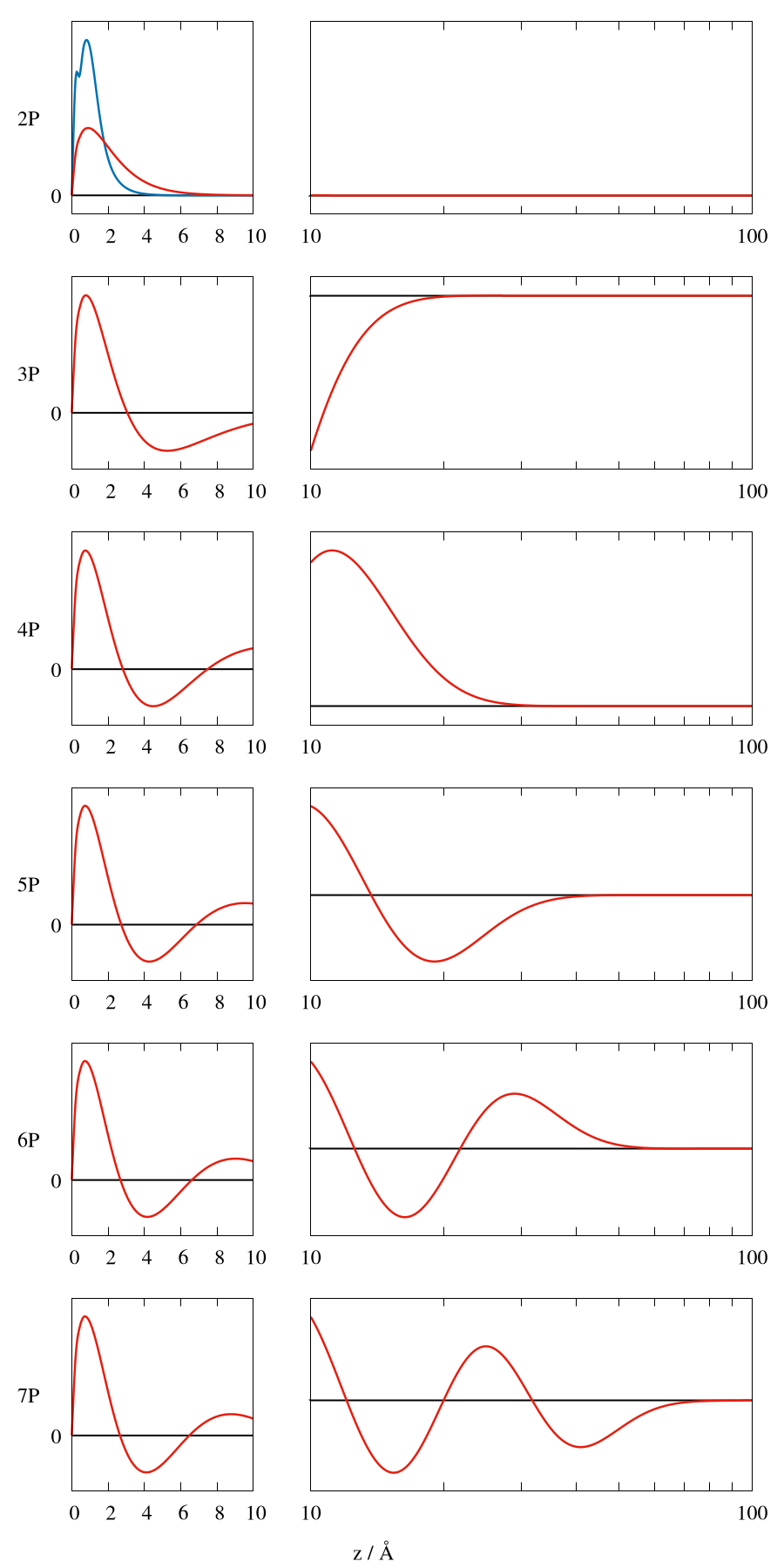}
\caption{\label{FigP}Red: Rydberg P$_z$ orbitals of the $^2$Li as
  1-dimensional cuts along the $z$~axis, and obtained as the RHF
  virtual orbitals of the $^1$Li$^+$ cation with negative orbital
  energy. Computed in the \mbox{AAQZ-ExTG-Ry7G} basis. Blue: Rydberg
  2P$_z$ orbital obtained in the \mbox{aug-ANO-pVQZ} basis.}
\end{figure}

\begin{figure}[htb]
\includegraphics[width=1.0\textwidth]{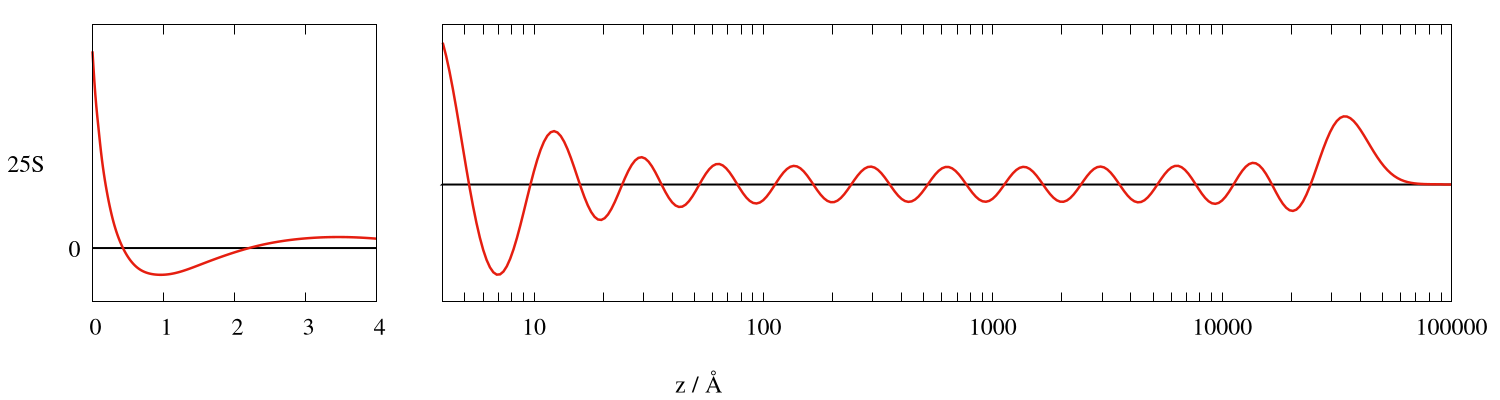}
\caption{\label{Fig25S}Rydberg 25S orbital of the $^2$Li as a
  1-dimensional cut along the $z$~axis, and obtained as the RHF
  virtual orbital of the $^1$Li$^+$ cation with negative orbital
  energy. Computed in the 2023Sm basis.}
\end{figure}

Figures \ref{FigS} and \ref{FigP} show the S and P Rydberg orbitals
computed in the \mbox{AAQZ-ExTG-Ry7G} basis.  With each excitation
level the number of nodes increases by one and spatially the orbitals
also reach one more step further from the nucleus on the logarithmic
scale of the $z$ axis. For the highest computed 7S and 7P orbitals
there are six and five nodes, respectively, reaching tens of {\AA} far
from the nucleus. The wave-function oscillations and the nodes are
very regular in the logarithmic scale of the $z$ axis and could
actually define the size of the Rydberg excited atom. One can also
notice that the distances between the inner-most nodes tend to be
wider for P than for the S orbitals. This is a trend for the higher
angular momenta, too (see the SI), where the inner nodes of the S
states are as narrow as 2 {\AA} and gradually widen to about 15 {\AA}
for the G symmetry orbitals.

Figure \ref{FigP} also depicts the 2P orbital computed with a bare
\mbox{aug-ANO-pVQZ} basis, which is the only Rydberg orbital the basis
is able to model (except for the valence 2S orbital). The standard
basis apparently failed to provide smooth orbital curve, and despite
being augmented, it exhibits too limited range --- dissipating the 2P
orbital at about 4 {\AA} compared to the actual 7 {\AA} resulting from
the \mbox{AAQZ-ExTG-Ry7G} basis. This example illustrates the
importance of the tailored Rydberg basis generation for studying
diffuse systems. Considering that only a few more optimized functions
are necessary for an accurate reproduction of the Rydberg orbitals
(the minimal state-selective basis \mbox{AAQZ-ExTG-Ry3P} adds just two
S and two P functions to the \mbox{aug-ANO-pVQZ}), the methodology may
thus promise a substantial impact on {\it ab~initio} quantum chemistry
of Rydberg or anionic systems with negligible additional computational
complexity.

The SI contains plots of all the Rydberg orbitals achieved by the
2023Sm basis, which have not yet been presented before. All the
orbitals computed by the \mbox{AAQZ-ExTG-Ry7G} basis ({\it
  \mbox{i. e.}} up to $n=7$) are visually indistinguishable from those
computed with the 2023Sm basis. The higher excited orbitals have also
regular nodal structure in accord to the above described
observations. Nevertheless, the higher D orbitals have noticeable
flaws, which match with the discussed poor performance of the high D
Rydberg states computed with the 2023Sm basis.~\cite{2023Sm} Figure
\ref{Fig25S} shows the highest 25S Rydberg orbital achieved by the
2023Sm basis with a clear and regular nodal structure along the
logarithmic $z$ axis. The wave-function lobe behind the last 24th node
reaches almost up to $100,000$~{\AA}, {\it \mbox{i. e.}} 10 $\mu$m far
from the nucleus. This means that the diameter of this excited atom is
already comparable to the scale of bacteria. That such a large atomic
structure could be described {\it \mbox{ab initio}} with a relatively
small contracted Gaussian basis, only emphasizes the significance of
the developed methodology.

%%%%%%%%%%%%%%%%%%%%%%%%%%%%%%%%%%%%%%%%%%%%%%%%%%%%%%%%%%%%%
\section{Conclusions}
\label{Conclusions}

A new Gaussian basis set tailored for an accurate description of
$^2$Li Rydberg excited states up to $n=7$ and for S, P, D, F, and G
angular momentum symmetry has been developed. In contrast to the
earlier published lithium Rydberg basis~\cite{2023Sm}, the new basis
set is very small, both in the number of utilized Gaussian primitives
and the number of contracted basis functions, especially for the
state-selective calculations, which need only two S and two P
optimized Rydberg functions added to a standard basis set. Even with
such a reduced size the new basis still preserves the accuracy of the
correlated excitation energies and the ionization potential higher
than $10^{-2}$~eV. Instead of using the even tempered Gaussians, the
new basis has been parameterized with the exponentially tempered
Gaussian primitives, which made the basis more flexible and less prone
to near linear dependence issues during the optimization process.

At the EOM-CCSD level of theory, the new basis also exhibits
comparable quality of the S Rydberg states as the universal Gaussian
primitive set~\cite{1989KaBaJu}, but requiring notably smaller number
of optimized basis functions. For higher angular momenta, the new
basis is undoubtedly superior to the universal basis set. The
comparison thus confirms that the exponentially tempered Gaussian
primitives are highly appropriate for the description of the Rydberg
orbitals and also suggests that performing the tailored basis set
optimization is more favourable than using universal basis sets.

By plotting one-dimensional wave-function cuts of the resulting
Rydberg orbitals, a regular nodal structure has been observed on a
logarithmic scale of the atomic radius, reaching higher tens of {\AA}
far from the nucleus for the $n = 7$ states. A plot of the 25S state,
the highest S state achieved by the earlier developed
basis~\cite{2023Sm}, revealed the regular nodal structure that reached
almost to $100,000$~{\AA}, {\it i.~e.}~$10\, \mu m$, far from the
nucleus --- a scale comparable to the size of bacteria. Considering
that an accurate description of an excited atom of such an enormous
dimension has been made possible with a basis composed of a relatively
small number of Gaussians, only underscores the undervalued
capabilities of the quickly dissipating Gaussian-type basis sets for
the description of Rydberg states, and emphasizes the significance of
the developed basis set optimization methodology.

In summary, the relatively simple tailored basis set generation
methodology has been substantially improved and is now able to provide
a highly accurate description of diffuse lithium Rydberg states with
bare minimal Gaussian basis set requirements. It is a promising step
towards routine {\it ab initio} studies of more complex Rydberg-state
related systems, such as large atoms and polyatomic
molecules. Nevertheless, developing the basis set generation protocol
for such complex systems will still need more investigation.

%%%%%%%%%%%%%%%%%%%%%%%%%%%%%%%%%%%%%%%%%%%%%%%%%%%%%%%%%%%%%
\section{Disclosure Statement}
\label{Disclosure}

No potential conflict of interest was reported by the authors.

%%%%%%%%%%%%%%%%%%%%%%%%%%%%%%%%%%%%%%%%%%%%%%%%%%%%%%%%%%%%%
\section{Acknowledgment}
\label{Ack}

The authors gratefully acknowledge the financial support by the
SVV–2023–260716 grant from Charles University.  Computational
resources were provided by the e-INFRA CZ project (ID:90254),
supported by the Ministry of Education, Youth and Sports of the Czech
Republic.

%%%%%%%%%%%%%%%%%%%%%%%%%%%%%%%%%%%%%%%%%%%%%%%%%%%%%%%%%%%%%
% Create the reference section using BibTeX:
\bibliography{LiRy}

\end{document}